\documentclass[12pt]{article}
\usepackage{amssymb}
\usepackage{amsmath}
\usepackage{epsfig}

\begin{document}

\title{The Ups and Downs of Cyclic Universes}
\author{T. Clifton\thanks{
e-mail: T.Clifton@cantab.net } \\
{\small {\textit{Department of Physics, Stanford University, CA 94305, USA}} 
\vspace{1mm}}\\
and \vspace{1mm}\\
John D. Barrow\thanks{
e-mail: J.D.Barrow@damtp.cam.ac.uk}\\
{\small {\textit{Department of Applied Mathematics and Theoretical Physics,}}
}\\
{\small {\textit{University of Cambridge, Wilberforce Rd., }}}\\
{\small {\textit{Cambridge CB3 9LN, UK}}}}
\date{{\normalsize {\today}}}
\maketitle

\begin{abstract}
We investigate homogeneous and isotropic oscillating cosmologies with
multiple fluid components. Transfer of energy between these
fluids is included in order to model the effects of non-equilibrium behavior on closed
universes. We find exact solutions which display a range of new behaviors
for the expansion scale factor. Detailed examples are studied for the
exchange of energy from dust or scalar field into radiation.  We show that,
contrary to expectation, it is unlikely that such models can offer a
physically viable solution to the flatness problem.
\end{abstract}

\section{Introduction}

In this paper we consider a wide range of homogeneous and isotropic
cosmological models containing two fluids which are able to
exchange energies and manifest non-equilibrium behavior. Recently
\cite{Cli06}, we studied a class of such examples in the dynamical context
of a flat Friedmann universe. A simple ansatz was used to model energy
transfer between two fluids, which would be separately conserved perfect
fluids in the absence of the energy transfer. By means of this simple
formulation
we were able to find a single master equation whose range of solutions
allowed us to study the cases of massive particles decaying into radiation,
particle-antiparticle annihilation into radiation, the decay of classical
vacuum energy, and the formation and Hawking evaporation of black holes. In
this paper, we add some further ingredients: the effects of spatial curvature
are included in the cosmological dynamics and a wider range of energy
exchanges are considered. The most interesting situation allowed by the
introduction of spatial curvature is the possibility of an oscillating
closed universe. Following Tolman's \cite{tol} identification of the
important role that the second law of thermodynamics could play in the long 
term evolution of a cyclic universe in which entropy production takes place,
there have been further detailed studies of how thermodynamics leads to
differences in the cycle-to-cycle evolution. Landsberg and Park \cite{land}
studied the
approach to flatness created by the growing cycles; Barrow and
D\c{a}browski \cite{bdab} carried out a detailed study of the evolution of
anisotropies and black holes, and also showed that if a positive
cosmological constant is present then Tolman's growing oscillations will
eventually be ended -- replaced by unending expansion towards the de Sitter
space-time. However, these studies did not attempt to model the action of
energy exchanges between matter and radiation; they simply injected entropy
(in the form of radiation) into the universe at each moment of bounce. A
more realistic modelling of energy exchange was introduced by Barrow,
Kimberly, and Magueijo \cite{bmk}, who included energy transfer between
scalar fields and radiation in some cosmological models arising in general
relativity, Brans-Dicke theory, and theories in which variations in the
fine structure `constant' are driven by appropriately coupled scalar fields 
\cite{bmag}. This
revealed the interesting feature that oscillating universes, in which time
variations in the gravitational `constant' and the fine structure `constant'
can occur, saw those variations continuing in an almost monotonic fashion
from
cycle to cycle. Unlike the expansion scale factor, these varying `constants'
did not oscillate.  A further variant was considered by the present authors, in ref 
\cite{BCgrav}, where they considered energy exchange between radiation and
the Brans-Dicke scalar
field so that energy was drained from the gravitational `constant' as the
universe expanded. Here, we extend these investigations and enlarge the
gallery of possible cosmological evolutions of the scale factor that can
occur in oscillating closed universes in general relativity.

\section{Cosmological Models}

Consider a spatially homogeneous and isotropic Friedmann universe, with
expansion scale factor $a(t),$ containing two fluids with equations of state
\begin{eqnarray*}
p &=&(\gamma -1)\rho , \\
p_{1} &=&(\Gamma -1)\rho _{1},
\end{eqnarray*}
where the $\gamma $ and $\Gamma $ are constants, $p$ and $p_{1}$ are the
fluid pressures and $\rho $ and $\rho _{1}$ are their densities. The
Friedmann equation is then given by
\begin{equation}
H^{2}=\rho +\rho _{1}-\frac{k}{a^{2}},  \label{1}
\end{equation}
where $8\pi G/3\equiv 1$, $H=\dot{a}/a$ is the Hubble expansion rate of
the universe, and $t$ is comoving proper time. We
write the evolution equations for these two fluids as
\begin{eqnarray}
\dot{\rho}+3H\gamma \rho &=&s,  \label{2} \\
\dot{\rho}_{1}+3H\Gamma \rho _{1} &=&-s  \label{3}
\end{eqnarray}
where $s$ is a function parameterising any exchange of energy between them.

In general, $s$ could be any function of $H$, $\rho $, $\rho _{1},t$
and $a$.  In a previous work \cite{Cli06}, we considered the case where $s$ was
prescribed by
\begin{equation}
s=-\alpha H\rho +\beta H\rho _{1},  \label{old}
\end{equation}
with $\alpha $ and $\beta $ both constant, and $k=0$. This case can be used
to model the two-way exchange of energy between fluids in a number of
physically interesting situations involving particle decays, vacuum decay,
particle-antiparticle annihilation, or black hole formation and evaporation,
and is particularly appealing because it can be solved completely in terms
of simple functions. Here, we will consider more general formulations of
this problem, particularly those that arise when there is non-zero spatial
curvature ($k\neq 0$) and with different dependence of the energy exchange
parameter, $s$, on the other physical variables.

The inclusion of spatial curvature allows particular cases of special
interest, such as oscillating universes with entropy production, to be
investigated. In these models a universe of positive spatial curvature
repeatedly expands to a maximum and recollapses towards a big crunch. In the
absence of non-adiabatic processes, these cyclic universes will all have the
same maximum of expansion and total lifetime. However, it was first argued
by Tolman \cite{tol} that an increase of entropy at the moment of each
crunch-to-bang transition should create a growth in the scale of successive
maxima of expansion. In
this way, the universe may approach spatial flatness at asymptotically
late times, allowing a potential solution of the flatness problem.  These
growing oscillations will continue until such time as any positive
cosmological constant, or quintessence field that violates the strong energy
condition, dominates the expansion, after which the oscillations will cease,
as first shown in ref \cite{bdab}.  If the entropy production occurs in relatively
small steps then the final state should also be one in which the energy
density in the quintessence field is just slightly greater than that in the
cold dark matter fields.

\section{Cosmological Thermodynamics}

The exchange of energy and momentum between different components of a
universe containing multiple fluids requires us to reconsider the
thermodynamics of the universe, especially the effects of the second law of
thermodynamics. We write the fundamental law of thermodynamics as
\begin{equation}
Td\mathbf{S}=d(\rho V)+pdV=d\left[ (\rho +p)V\right] -Vdp  \label{fun}
\end{equation}
where $\mathbf{S}$ (not to be confused with $s$) is the entropy of the fluid
and $V\equiv a^{3}$ is the comoving volume of the universe. The
integrability condition
\begin{equation*}
\frac{\partial ^{2}\mathbf{S}}{\partial T\partial V}=\frac{\partial ^{2} 
\mathbf{S}}{\partial V\partial T}
\end{equation*}
then gives the relation
\begin{equation*}
dp=\frac{(\rho +p)}{T}dT
\end{equation*}
which can be substituted into equations (\ref{2}) and (\ref{fun}) to give
\begin{equation*}
\frac{d}{dt}\left[ \frac{(\rho +p)V}{T}\right] =\frac{Vs}{T}\qquad \text{and}%
\qquad d\mathbf{S}=d\left[ \frac{(\rho +p)V}{T}\right] .
\end{equation*}
We can then read off the expression for the density $\rho $ fluid
\begin{equation}
\dot{\mathbf{S}}=\frac{sV}{T}  \label{s}
\end{equation}
and similarly, for $\rho _{1}$,
\begin{equation}
\dot{\mathbf{S}}_{1}=-\frac{sV}{T_{1}}.  \label{s1}
\end{equation}
It can be seen directly from these expressions that allowing a transfer of
energy between different components of the universe ($s\neq 0$) results in
the entropy of those components no longer staying constant. In the limit
that this interaction disappears, $s\rightarrow 0$, the rate of change of
entropy also disappears and entropy is again conserved.
These expressions for the rate of change in entropy of the two fluids enable
us to assess the implications of the second law of thermodynamics. According
to the second law:
\begin{equation*}
0\leqslant \dot{\mathbf{S}}_{T}\equiv \dot{\mathbf{S}}+\dot{\mathbf{S}}%
_{1}=sV\left( \frac{1}{T}-\frac{1}{T_{1}}\right) .
\end{equation*}
This equation tells us that for energy to be transferred from $\rho $ to 
$\rho _{1}$ (i.e. $s<0$) we must have $T\geqslant T_{1}$ and that,
conversely, for energy to be transferred from $\rho _{1}$ to $\rho $ ($s>0$)
we must have $T_{1}\geqslant T$. In other words: energy is always
transferred from the hotter to the cooler fluid.

\section{Exact Solutions}

There are a number of exact solutions that can be found to the system of
equations (\ref{1}), (\ref{2}) and (\ref{3}). In this section we will
present methods of integrating these equations, the analysis of which will
follow in the subsequent sections of the paper.

\subsection{Solutions with $\mathbf{\protect\rho=\protect\rho(a)}$}

Taking the form of $s$ given by (\ref{old}) and making the definitions
\begin{eqnarray*}
A &\equiv &\alpha +\beta +3(\gamma +\Gamma ), \\
B &\equiv &\frac{3}{2}(\alpha \Gamma +\beta \gamma +3\gamma \Gamma ),
\end{eqnarray*}
the equations (\ref{2}) and (\ref{3}) can be put in the simple form
\begin{equation*}
\frac{\rho ^{\prime \prime }}{\rho }+A\frac{\rho ^{\prime }}{\rho }+2B=0,
\end{equation*}
where primes denote differentiation with respect to the new time coordinate,
\begin{equation*}
\eta \equiv \ln a.
\end{equation*}
The solution to this equation is then given by
\begin{eqnarray*}
\rho &=&c_{1}e^{\frac{1}{2}(-A-\sqrt{A^{2}-8B})\eta }+c_{2}e^{\frac{1}{2}%
(-A+ \sqrt{A^{2}-8B})\eta } \\
&=&c_{1}a^{\frac{1}{2}(-A-\sqrt{A^{2}-8B})}+c_{2}a^{\frac{1}{2}(-A+\sqrt{
A^{2}-8B})}
\end{eqnarray*}
where $c_{1}$ and $c_{2}$ are constants. Substitution into (\ref{3}) then
gives $\rho _{1}\propto \rho $ and so (\ref{1}) can be rewritten as
\begin{equation*}
H^{2}=c_{3}a^{\frac{1}{2}(-A-\sqrt{A^{2}-8B})}+c_{4}a^{\frac{1}{2}(-A+\sqrt{
A^{2}-8B})}-\frac{k}{a^{2}}.
\end{equation*}
It can be seen that this equation reduces to the solution found in ref {%
\cite{Cli06}} 
in the limit $k\rightarrow 0$, and that the inclusion of spatial
curvature does not add any new features to the form of $a$ that are not
already present in the standard evolution without interaction. Specifically,
it can be seen that in this case an oscillating universe with $k>0$ will
have the same maximum of expansion in each of its cycles.

\subsection{Solutions with $\mathbf{\protect\rho=\protect\rho(a, t)}$}

The Friedmann equation (\ref{1}) shows that, for any
functions $\rho =\rho (a)$ and $\rho _{1}=\rho _{1}(a)$, the evolution of the
universe shares the same property as the example above. Oscillating $k>0$
universes have constant amplitude. In order to find new behaviors, we will
consider a more general situation where we parameterise $\rho =\rho (a,t)$
and $\rho _{1}=\rho _{1}(a,t)$. This will allow new exact solutions to be
found with two interacting fluids, and non-zero spatial curvature.
Integrating equations (\ref{2}) and (\ref{3}), we obtain
\begin{equation}
\rho =\frac{\int sa^{3\gamma }dt}{a^{3\gamma }}\equiv \frac{m}{a^{3\gamma }}%
\qquad \qquad \text{and}\qquad \qquad \rho _{1}=\frac{-\int sa^{3\Gamma }dt}{
a^{3\Gamma }}\equiv \frac{m_{1}}{a^{3\Gamma }},  \label{m}
\end{equation}
which defines the quantities $m$ and $m_{1}$. In the absence of energy
exchange ($s=0$), these two quantities are constant and we have the same
form for $\rho $ and $\rho _{1}$ as in the usual case. As soon as energy
exchange is allowed, $m$ and $m_{1}$ become non-constant, corresponding to
energy being exchanged between the two fluids.

\subsubsection{A universe containing radiation and dust or massless scalar
field}

For the cases of universes containing radiation and dust, or radiation and a
scalar field, the field equations (\ref{1}), (\ref{2}) and (\ref{3}) can be
transformed into the equation of motion for a forced harmonic
oscillator. Such equations can be solved exactly using standard techniques.
These cases are of particular interest for modelling an oscillating universe
as they correspond to physical significant situations. It has often been
hypothesised that an oscillating universe containing these fluids should
allow for energy to be transferred, and entropy to be increased, at the
moment of crunch-to-bang. The case of exchange between radiation and a scalar
field is of particular interest as the kinetic parts of scalar fields may
dominate the earliest stages of the universe, and non-equilibrium behavior
can arise through slow or fast decays, as expected at the end of inflation
or other phase transitions involving these fields. Negative-energy scalar
fields are often used in oscillating universe models to produce a non-zero
minimum of expansion, and a detailed study of their effects, especially on
any `constant' of Nature which may vary in time was given in ref \cite{bmk}.

Differentiation of the Friedmann equation (\ref{1}), and some manipulation
and substitution using (\ref{2}) and (\ref{3}) allows us to write\ the
evolution equation for $H(t)$ as
\begin{equation*}
\dot{H}+\frac{3}{2}\Gamma H^{2}=\frac{3}{2}(\Gamma -\gamma )\rho +(1-\frac{3 
}{2}\Gamma )\frac{k}{a^{2}}.
\end{equation*}
Following the prescription given in ref \cite{jbobs} to turn
non-dissipative Friedmann universes into simple harmonic oscillators in
conformal time, we make a transformation to the conformal time coordinate $%
ad\tau =dt$  and define the new variable $b=a^{3\gamma -2}$, to obtain the
equation of motion of a forced harmonic oscillator:
\begin{equation}
b^{\prime \prime }+(2-3\gamma )^{2}kb=\frac{1}{2}(2-3\gamma )^{2}m,
\label{SHO}
\end{equation}
where primes here denote differentiation with respect to $\tau $. In
deriving this equation we have used the definition of $m$ given by (\ref{m}%
), and assumed the relation $3\Gamma +2=6\gamma $: This includes the
important cases of radiation and dust ($\Gamma =4/3$ and $\gamma =1$) and
radiation and scalar field ($\gamma =4/3$ and $\Gamma =2$), which this
section focuses upon. The well known solution to equation (\ref{SHO}), for
general $m$ and $k\neq 0$, is given by
\begin{eqnarray*}
b(\tau ) &=&c_{1}\sin \left( \sqrt{(2-3\gamma )^{2}k}\tau \right) +c_{2}\cos
\left( \sqrt{(2-3\gamma )^{2}k}\tau \right) \\
&&+\frac{\sqrt{(2-3\gamma )^{2}}}{2k}\sin \left( \sqrt{(2-3\gamma )^{2}k}%
\tau \right) \int m\cos \left( \sqrt{(2-3\gamma )^{2}k}\tau _{1}\right)
d\tau _{1} \\
&&-\frac{\sqrt{(2-3\gamma )^{2}}}{2k}\cos \left( \sqrt{(2-3\gamma )^{2}k}%
\tau \right) \int m\sin \left( \sqrt{(2-3\gamma )^{2}k}\tau _{1}\right)
d\tau _{1},
\end{eqnarray*}
where $c_{1}$ and $c_{2}$ are constants of integration. By specifying
a particular energy-exchange function, $s$, the value of $m$ can be found
from
equation (\ref{m}), and the integrals above can then be performed. Methods
of obtaining various forms of $m(t)$ are outlined in the appendix.

If we take, for example, the forms of $s$ to be
\begin{equation*}
s_{A}=s_{0}\rho a^{-1}\qquad \qquad \text{and}\qquad \qquad
s_{B}=s_{0}a^{-(1+3\gamma )},
\end{equation*}
where $s_{0}$ is a constant, then the corresponding functions $m$ are found
in the appendix to take the particularly simple forms
\begin{equation*}
m_{A}=m_{0}e^{s_{0}\tau }\qquad \qquad \text{and}\qquad \qquad
m_{B}=m_{0}+s_{0}\tau ,
\end{equation*}
where $m_{0}$ is a constant of integration. The solutions, for $k\neq 0$,
can then be written as
\begin{equation}
b_{A}=c_{1}\sin \left( \sqrt{(2-3\gamma )^{2}k}(\tau -\tau _{0})\right) +%
\frac{(2-3\gamma )^{2}m_{0}e^{s_{0}\tau }}{2((2-3\gamma )^{2}k+s_{0}^{2})}
\label{bA}
\end{equation}
and
\begin{equation}
b_{B}=\bar{c}_{1}\sin \left( \sqrt{(2-3\gamma )^{2}k}(\tau -\bar{\tau}%
_{0})\right) +\frac{(m_{0}+s_{0}\tau )}{2k}  \label{bB}
\end{equation}
where $c_{1}$, $\bar{c}_{1}$, $\tau _{0}$ and $\bar{\tau}_{0}$ are
constants, and subscripts $A$ and $B$ denote solutions corresponding to
$s_{A}$ and $s_{B}$, respectively. The behavior of these solutions will be
investigated in the next section. The case of no energy being exchanged can
be read off from equation (\ref{bB}) as the sub-class of solutions with
$s_{0}=0$.

\subsubsection{A universe containing radiation and non-zero vacuum energy}

A second mechanism for obtaining an oscillating universe is a negative
vacuum energy. Due to the lack of dissipation of vacuum energy as the
universe expands, this fluid always comes to dominate at late-times (if the
universe survives that long, and in the absence of phantoms). Negative
vacuum energy then plays the role that was previously played by positive
spatial curvature, and causes the universe to collapse at late-times as
anti-de Sitter behavior is approached.

We will now proceed to find solutions for a universe containing radiation
and non-zero vacuum energy ($\rho =-p$). Again, the Friedmann equations can
be cast into the form of the equation of motion for a harmonic oscillator,
but now the forcing term will be constant and the analogue of the mass will
be non-constant.

Taking $\Gamma =4/3$, $\gamma =0$ and defining the new variable $\alpha
\equiv a^{2},$ allows the equations (\ref{1}), (\ref{2}) and (\ref{3}) to be
reduced to the form
\begin{equation}
\ddot{\alpha}-\frac{4\Lambda }{3}\alpha =-2k,  \label{SHO2}
\end{equation}
where use has been made of (\ref{m}), and $\Lambda \equiv 3m=3\rho $ has
been defined in analogy to the usual notation of the cosmological constant.
(The reader should bear in mind that $\Lambda $ is \textit{not} a constant
here, as energy is being exchanged between it and the radiation fluid). This
equation can be solved once $s$ has been specified, and a solution obtained
for $\Lambda $.

If we now take, for example, the energy-exchange parameters
\begin{equation*}
s_{C}=s_{0}\Lambda \qquad \qquad \text{and}\qquad \qquad s_{D}=s_{0}
\end{equation*}
then, from the appendix, we obtain the functions
\begin{equation*}
\Lambda _{C}=\Lambda _{0}e^{3s_{0}t}\qquad \qquad \text{and}\qquad \qquad
\Lambda _{D}=\Lambda _{0}+3s_{0}t,
\end{equation*}
where $s_{0}$ and $\Lambda _{0}$ are constants. The equation (\ref{SHO2})
can now be solved for these $s$ and $m$, to give
\begin{equation}
\alpha _{C}=c_{3}I_{0}\left( x\right) +c_{4}K_{0}\left( x\right) -\frac{2k}{
3s_{0}}I_{0}\left( x\right) \int K_{0}\left( x_{1}\right) dt_{1}+\frac{2k}{
3s_{0}}K_{0}\left( x\right) \int I_{0}\left( x_{1}\right) dt_{1}  \label{aA}
\end{equation}
and
\begin{equation}
\alpha _{D}=\bar{c}_{3}A_{i}\left( y\right) +\bar{c}_{4}B_{i}\left( y\right)
+ks_{0}\sqrt[3]{\frac{\pi ^{3}}{12s_{0}^{4}}}A_{i}\left( y\right) \int
B_{i}\left( y_{1}\right) dt_{1}-ks_{0}\sqrt[3]{\frac{\pi ^{3}}{12s_{0}^{4}}}%
B_{i}\left( y\right) \int A_{i}\left( y_{1}\right) dt_{1}  \label{aB}
\end{equation}
where $c_{3}$, $\bar{c}_{3}$, $c_{4}$ and $\bar{c}_{4}$ are constants,
$I_{0} $ and $K_{0}$ are Bessel functions, $A_{i}$ and $B_{i}$ are Airy
functions and $x$ and $y$ are defined by
\begin{equation*}
x=\sqrt{\frac{16|\Lambda _{0}|}{27s_{0}^{2}}}e^{3s_{0}t/2}\qquad \qquad 
\text{and}\qquad \qquad y=\sqrt[3]{\frac{4}{27s_{0}^{2}}}(\Lambda
_{0}+3s_{0}t).
\end{equation*}
The form of $\alpha $ given in (\ref{aA}) corresponds to $\Lambda _{0}>0$;
making the substitutions $K_{0}\rightarrow \pi J_{0}$ and
$I_{0}\rightarrow Y_{0}$ gives the solution for $\Lambda _{0}<0$. For
$k=0$ the unsightly 
terms involving integrals of Bessel and Airy functions in (\ref{aA}) and (\ref{aB}) vanish.

In the absence of any energy exchange, $\Lambda $ is constant, and the
solution to equation (\ref{SHO2}) is
\begin{equation*}
\alpha =c_{5}\sin \left\{ \sqrt{\frac{-4\Lambda }{3}}(t-t_{1})\right\} +%
\frac{3k}{2\Lambda }
\end{equation*}
where $c_{5}$ and $t_{1}$ are constants. In the next section we will proceed
to investigate the physical behavior of these solutions.

\section{Behavior of the Cosmological Solutions}

Having found exact solutions for oscillating universes with energy exchange,
in the previous section, we will now investigate their behavior. The
oscillatory nature of these solutions is created either by positive spatial
curvature, or a negative vacuum energy, either of which will cause a maximum
of expansion, after which the universe collapses. By introducing energy
exchange between the fluids in these models, we allow for the possibility of
different cycles having different expansion maxima.

\subsection{A closed universe containing radiation and scalar field}

From equations (\ref{bA}) and (\ref{bB}) above, we see that the scale-factor
for a universe containing radiation ($\gamma=4/3$) and scalar field
($\Gamma=2$) with positive spatial curvature ($k=1$) can be written as
\begin{equation}  \label{bA2}
a^2_A = c_1 \sin \left\{ 2 (\tau-\tau_0) \right\} + e^{s_0 (\tau-\tau_1)}
\end{equation}
when $s=s_0 \rho a^{-1}$, and as
\begin{equation}  \label{bB2}
a^2_B = \bar{c}_1 \sin \left\{ 2 (\tau-\bar{\tau}_0) \right\} + \frac{1}{2}
s_0(\tau-\bar{\tau}_1)
\end{equation}
when $s=s_{0}a^{-5}$. Here we have absorbed some of the constants of (\ref%
{bA}) and (\ref{bB}) into the new constants $\tau _{1}$ and $\bar{\tau}_{1}$.

\begin{figure}[ht]
\center \epsfig{file=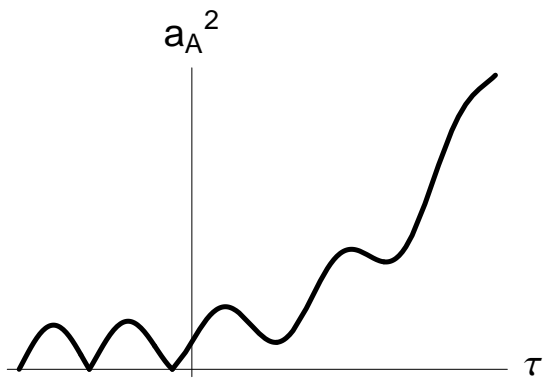,height=5cm}
\caption{{{\textit{The evolution of the scale factor in a closed universe
with radiation and scalar field exchanging energy, as prescribed in equation
(\protect\ref{bA2}) when $0<s\propto \protect\rho a^{-1}$.}}}}
\label{bAf}
\end{figure}

\begin{figure}[ht]
\center \epsfig{file=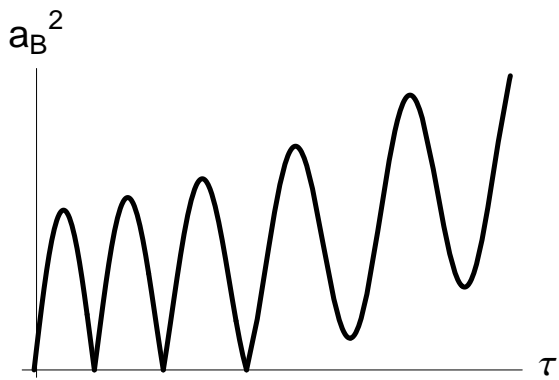,height=5cm}
\caption{{{\textit{The evolution of the scale factor in a closed universe
with radiation and scalar field exchanging energy, as prescribed in equation
(\protect\ref{bB2}) when $0< s\propto a^{-5}$.}}}}
\label{bBf}
\end{figure}

The two functions (\ref{bA2}) and (\ref{bB2}) above have the undesirable
feature of allowing $a^{2}$ to be both positive and negative, for any set of
the constants $c_{1}$, $\tau _{0}$ and $\tau _{1}$. However, only two of
these three will be fixed by specifying the energies of the two fluids at
the moment of collapse; the third is a free constant which can be used to
match together different cycles, at the moment of crunch-to-bang. In this
way the scale-factor can be made real and positive semi-definite, throughout
its evolution. Figures \ref{bAf} and \ref{bBf} show the evolution of these
universes after such a matching.

It can be seen from figures \ref{bAf} and \ref{bBf} that, as time progresses,
the expansion maximum of each cycle increases. This is because we chose $s_{0}>0$ to
construct these plots, corresponding to energy being transferred from the
scalar field to radiation. This causes successive maxima of expansion to
increase in magnitude, due to the extra radiation. This is in accord with
Tolman's original model \cite{tol}. However, as well as the maxima
increasing from cycle to cycle, we also see that after a certain number of
cycles there exist real non-zero minima of expansion, instead of collapse to
a singularity. After this point the evolution of the universe is
non-singular. It can clearly be see that these minima of expansion also
increase in magnitude as $\tau $ increases. This is an effect that has not
been recognised in earlier discussions of oscillating universes subject to
the second law of thermodynamics. We interpret this behavior as being due to
the scalar field
transferring so much energy to the radiation field that its own energy
density becomes negative, allowing a non-zero minimum of expansion. Negative
energy scalar fields are often used in this way to create oscillating
universe models without singularities; however, we can include a cut-off to
the energy transfer that would prevent this from occurring.

In addition to this common behavior, significant differences can also be
seen to occur between figures \ref{bAf} and \ref{bBf} at both early and
late-times. Such differences in behavior clearly illustrate that the
evolution of these universes is sensitive to the particular form of energy
exchange that has been chosen.

At late-times, figure \ref{bAf} can be seen to exhibit a runaway behavior.
As the energy density of the radiation in this universe increases, so does
the rate of energy transfer (as it is proportional to $\rho $). This results
in the endless transfer of energy, at ever increasing rates, from the
unbounded negative energy scalar to the radiation. In fact, after a
sufficient number of cycles the runaway transfer of energy eventually
becomes so great that it overwhelms the oscillatory nature of the solutions,
after which these universes expand eternally. The late-time behavior of
figure \ref{bBf} also allows an endless transfer of energy from the negative
energy scalar to radiation. However, now the energy transfer is no longer
increased by the ever-increasing energy density in radiation, $\rho $, and
so does not display the runaway behavior seen in figure \ref{bAf}. The
increase of each expansion maximum and minimum happens at a steadier rate,
and at no point does the energy exchange become rapid enough to overwhelm
the oscillatory nature of the solutions.

At early-times, these solutions exhibit cycles which are separated by
catastrophic collapse to a singularity. As time is run backwards, the
maximum of each cycle decreases, as the energy of the radiation decreases.
Figure \ref{bAf} shows approach to a regime where each cycle is of the same
amplitude, in the limit, as $\rho \rightarrow 0$ and hence
$s_{A}\rightarrow 0 $. Expansion is then due to the scalar field, until it is inevitably
halted by the positive spatial curvature. By contrast, in figure \ref{bBf}
we see that the expansion maximum and the total duration of each cycle
continues to decrease, as time is run backwards. This behavior occurs
because the energy density of radiation is no longer bounded from below, and
continues to decrease until it becomes negative. After this point, the
radiation contributes to the onset of collapse, increasingly so as its
energy density becomes more negative. Of course, negative energy radiation
may not be particularly realistic, and we can include a
cutoff in the energy transfer function to prevent this from occurring. Such
a cutoff will eventually result in cycles of constant amplitude, when energy
is no longer
being transferred. It might also be avoided by picking forms for $s$ which
are more physically motivated, and may have a more complicated evolution in time,
rather than the simple choices we have made
for illustrative purposes.

The case of energy transfer from radiation to the scalar field can be
pictured using figures \ref{bAf} and \ref{bBf} by reversing the direction of
time.

\subsection{A closed universe containing radiation and dust}

From equations (\ref{bA}) and (\ref{bB}) we now see that a closed universe
containing radiation ($\Gamma=4/3$) and pressureless dust ($\gamma=1$) has
scale-factor
\begin{equation}  \label{bA3}
a_A = \hat{c}_1 \sin ( \tau-\hat{\tau}_0 ) + e^{s_0 (\tau-\hat{\tau}_1)}
\end{equation}
when $s=s_0 \rho a^{-1}$, and
\begin{equation}  \label{bB3}
a_B = \tilde{c}_1 \sin ( \tau-\tilde{\tau}_0 ) + \frac{1}{2} s_0(\tau-\tilde{
\tau}_1)
\end{equation}
when $s=s_0 a^{-4}$. Again, we have absorbed some of the constants of
(\ref{bA}) and (\ref{bB}) into the new constants $\hat{\tau}_1$ and $\tilde{\tau}_1$.

The behavior of the scale-factor is qualitatively the same as that shown in
figures \ref{bAf} and \ref{bBf} (with the axis label $a^{2}$ being replaced
by $a$). The main difference here is that $s_{0}>0$ (the direction of time
going from left to right in the figures above) now corresponds to energy
being transferred from radiation to dust. This results in the subsequent
maxima and minima of expansion increasing as the energy density of the dust
increases. For the more realistic situation of energy going from dust to
radiation, one should follow the graphs backwards, from right to left. Real
non-zero minima of expansion are taken here to arise from the energy density
in the radiation field taking a negative value. Again, a cut-off in energy
transfer may be needed to prevent this from occurring.

\subsection{A flat universe containing radiation and vacuum energy}

\begin{figure}[ht]
\center \epsfig{file=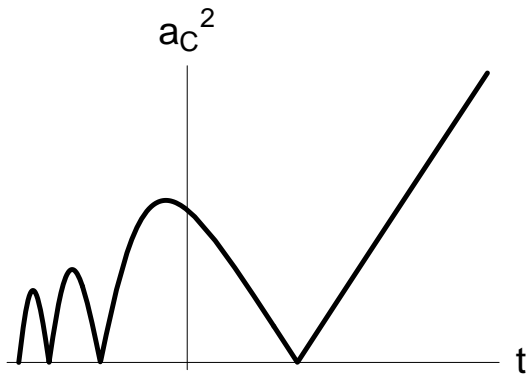,height=5cm}
\caption{{{\textit{The evolution of the scale factor in a flat universe with
radiation and vacuum exchanging energy, as prescribed in equation (  \protect
\ref{aA2}) when $0< s \propto \Lambda$.}}}}
\label{3f}
\end{figure}

\begin{figure}[ht]
\center \epsfig{file=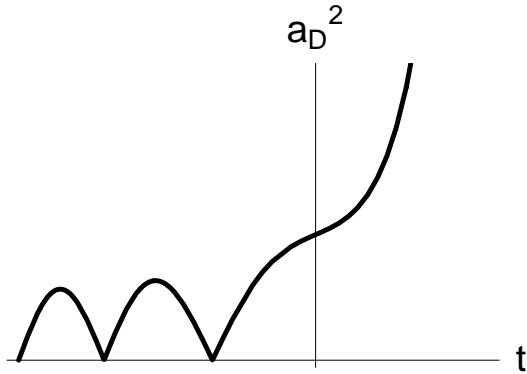,height=5cm}
\caption{{{\textit{The evolution of the scale factor in a flat universe with
radiation and vacuum exchanging energy, as prescribed in equation (  \protect
\ref{aB2}) when $0< s=$constant.}}}}
\label{4}
\end{figure}

From (\ref{aA}) and (\ref{aB}) it can be seen that a spatially flat universe
containing radiation ($\Gamma=4/3$) and negative vacuum energy ($\gamma=0$)
evolves as
\begin{equation}  \label{aA2}
a^2_C = c_2 Y_0 \left(x \right)+ c_3 \pi J_0 \left(x \right)
\end{equation}
when $s = s_0 \Lambda$, and as
\begin{equation}  \label{aB2}
a^2_D = \bar{c}_2 A_i \left(y \right)+ \bar{c}_3 B_i \left(y \right)
\end{equation}
when $s=s_0$, where $x$ and $y$ are the same as before. The form of these
solutions are shown in figures \ref{3f} and \ref{4}.

Again, these solutions will take both positive and negative values for any
set of $c_{2}$, $c_{3}$ and $\Lambda _{0}$, and likewise, solutions can be
matched at the moment of crunch-to-bang to ensure that $a$ remains real and
positive. This is done in figures \ref{3f} and \ref{4}.

Both of figures \ref{3f} and \ref{4} show universes which are initially
oscillating, with scale factor oscillations of increasing amplitude that
eventually end in an asymptotic period of continual expansion. This is due
to the choice $s>0$, which corresponds to energy being transferred from the
radiation to the negative vacuum energy. Once again, the contrary case of
energy transfer from vacuum energy to radiation can be considered by
reversing the direction of time in these plots.

In figure \ref{3f} we have the energy density of the vacuum evolving as 
$\Lambda =-|\Lambda _{0}|e^{-3|s_{0}|t}$, so that as $t\rightarrow \infty $
the energy density of the vacuum goes to zero. This has the effect of
leaving a radiation-dominated universe ($a^{2}\propto t$) in the late-time
limit, as can clearly be seen from the plot. The exchange of energy then
becomes unimportant as subsequent evolution continues in the standard way,
with the vacuum energy density having been forced to zero. At earlier times,
however, the influence of the negative vacuum energy becomes increasingly
important, resulting in oscillations of ever-decreasing amplitude. In
contrast to some of the previous cases, the energy density of the radiation
is always positive in this scenario, because increasing the magnitude of the
negative vacuum energy results in an increase in the positive energy density
of the radiation.

In figure \ref{4}, the transfer of energy from the vacuum to radiation
occurs at a constant rate, resulting in an evolution of the vacuum
energy density as $\Lambda =-|\Lambda |+3s_{0}t$. For $s_{0}>0$, energy is
transferred from the radiation to the vacuum, as in the previous example.
Now, however, the energy density of the vacuum is allowed to become positive
at sufficiently large $t$. This results in a period of rapidly accelerating
expansion as $t\rightarrow \infty $: not only is the evolution of this
universe dominated by a positive cosmological constant, but the value of
this \textquotedblleft constant\textquotedblright\ is itself increasing with
time. The early-time behavior is qualitatively similar in this example to
the previous one: as $t$ becomes increasingly negative, oscillations are
increasingly damped. Once again, we are guaranteed a positive energy density
of radiation, via the same mechanism as before. The principal difference
here is that the decrease in amplitude of oscillations, as $t$ decreases, is
slower than before because the transfer of energy is now constant, and not
proportional to $\Lambda $.

Inclusion of non-zero spatial curvature in these models results in a more
elaborate form of $a(t)$, as given in equations (\ref{aA}) and (\ref{aB}).
However, although more difficult to express in a concise analytic form, the
behavior of these solutions can be simply understood. In limits where
radiation previously dominated the evolution (such as the
$t\rightarrow \infty $ limit of figure \ref{3f}) spatial curvature now eventually
dominates. Positive curvature causes a maximum of expansion, and a
subsequent collapse, while negative curvature causes evolution towards a
Milne universe, $a\propto t$. In limits where the vacuum energy dominates
expansion, the spatial curvature has a negligible effect, unless it is large
enough to cause collapse before vacuum domination can occur.

\section{Discussion}

We have investigated the behavior of spatially homogeneous and isotropic
oscillating universes when non-equilibrium behavior is present. We have
found exact solutions for universes containing multiple fluids and non-zero
spatial curvature. These fluids include radiation and dust, scalar field, or
vacuum energy.

In the usual treatment, cosmologies with more than one fluid component are
assumed to evolve with these fluids staying non-interacting. The
energy of each fluid is then separately conserved, and the evolution of the
scale factor can be straightforwardly obtained by solving the Friedmann
equation. Here, we have considered a more general situation in which the
different fluid components of the universe are allowed to interact. Total
energy is still conserved in these models, but we have allowed energy to be
transferred between the different components. This situation has often been
discussed in the literature, with cosmological models in which
entropy increases only at each moment of crunch-to-bang. Here, we have
formulated
the problem consistently and found exact solutions in which entropy
production occurs realistically and we have used them to
model a number of different possibilities.

Firstly, we considered universes containing radiation and dust and universes
containing radiation and scalar field. In these models the oscillatory
nature of the solutions is due to positive spatial curvature. By allowing
interactions between these components, we find oscillating solutions where
the amplitude of successive cycles is allowed to vary. The case of radiation
and dust is of obvious interest as these components are required for the
nucleosynthesis of the light elements and formation of large-scale structure
in the standard cold dark matter model. The case of radiation and scalar
field is also of interest as scalar fields, if any exist in nature, will be
influential in the early evolution of the universe.
We find that the evolution of these universes is highly dependent upon the
exact form of interaction between the fluids. For some interactions, such as
those of the form considered in \cite{Cli06}, oscillating universe models
are found to progress with each of their cycles having identical
amplitude. For other interactions, it is found that energy transfer
leads to cycles of varying amplitude. Again, the precise from of this
variation is heavily dependent on the precise form of interaction. We find
general methods for solving the Friedmann equations, and investigate
explicitly two particularly simple models. It has often been conjectured
that oscillating anti-damped universes of growing amplitude may offer a
solution to the flatness
problem, without the need for inflation. By allowing the amplitude of each
cycle to increase monotonically, it is suggested \cite{land, bdab} that
eventually the
universe will end up being pushed closer and closer to spatial flatness,
with each successive cycle being longer and longer lived. Our toy models
show us that this may not be a particularly realistic expectation, even if
there is no cosmological constant. We find that the increase in amplitude of
the oscillations is halted in cases where the energy density of fluids is
bounded from below: there is simply not enough energy to allow the cycles
to become indefinitely large and long lived. In cases where the energy
densities are not bounded from below, such as are often considered with
scalar field models, we also find problems. When the energy density of these
components become negative the universe experiences a non-zero minimum of
expansion, instead of a crunch to singularity. This behavior is well known,
and is often used to model non-singular oscillating cosmologies. However, we
find that the continual transfer of energy required to create cycles of
indefinite size and duration results in subsequent increases in the non-zero
minimum of expansion. Thus, by sourcing energy from this negative energy
scalar field it becomes more negative, and increases in the maximum of
expansion are accompanied by increases in the minimum of expansion. This
behavior does not appear to be consistent with a physically viable cosmology
-- that is, one which starts small and lives long enough for matter
domination
to occur.

In addition to these models, we also consider universes containing radiation
and non-zero vacuum energy. The oscillatory nature of these solutions is
either produced by positive spatial curvature, or by a negative vacuum
energy.  For spatial curvature to have any non-negligible effect on these cosmologies
it must be strong enough to dominate the evolution of the universe before
the effects of vacuum energy become significant, otherwise vacuum effects
dominate. Again, we provide a general prescription for solving this problem
for arbitrary functions parameterising the exchange of energy. As before, we
give  two particularly simple toy models explicitly. In these simple models
we neglect spatial curvature and allow a negative vacuum energy to collapse
the universe. We find that if the vacuum energy is allowed to become
positive then it quickly comes to dominate, and accelerating expansion
rapidly ensues. On the other hand, if the vacuum energy becomes increasingly
negative then the amplitude and duration of cycles is found to decrease as
the
vacuum energy comes to dominate, and collapses the universe increasingly
rapidly. Finally, in examples where the exchange of energy causes the vacuum
energy to approach zero, we find that the subsequent evolution of the
universe progresses as a standard radiation-dominated Friedmann
universe.  Including the effects of spatial curvature complicates the precise analytic
form of the solutions, but otherwise acts in the expected way.

\vspace{20pt}

\leftline{\bf \Large Acknowledgements}

\vspace{10pt}

TC is supported by a Lindemann Fellowship.

\vspace{20pt}

\leftline{\bf \Large Appendix: From $\mathbf{s(a,t)}$ to
  $\mathbf{m(t)}$}

\vspace{10pt}

It can be seen from (\ref{2}) and (\ref{3}) that we can obtain any
invertible function $m=m(\tau )$ by choosing the energy exchange parameter
to take a form
\begin{equation*}
s=\left[ a^{3\gamma }f^{\prime }(\rho a^{3\gamma })g(a)\right] ^{-1},
\end{equation*}
where $f$ is the inverse function of $m=m(t)$ and $g(a)dT\equiv dt$ defines
the time coordinate $\tau $. Substituting this into (\ref{2}) gives
\begin{equation*}
f^{\prime }(\rho a^{3\gamma })\frac{d(\rho a^{3\gamma })}{d\tau }=1,
\end{equation*}
which integrates to
\begin{equation*}
f(\rho a^{3\gamma })=\tau -\tau _{0}
\end{equation*}
or
\begin{equation*}
\rho a^{3\gamma }=m(\tau -\tau _{0})=f^{-1}(\tau -\tau _{0}),
\end{equation*}
where $\tau _{0}$ is a constant of integration, which can be trivially
absorbed into a redefinition of the time coordinate $\tau \rightarrow
\tau +\tau _{0}$. This gives us a prescription for the form of $s(\rho ,a)$
required to derive any invertible function $m=m(\tau )$.

We will often be interested in the simple forms
\begin{equation}
m_{1}=m_{0}e^{s_{0}\tau }\qquad \qquad \text{and}\qquad \qquad
m_{2}=m_{0}+s_{0}\tau  \label{m2}
\end{equation}
where $m_{0}$ and $s_{0}$ are constants. It can now be seen from the above
that these forms for $m$ correspond to
\begin{equation}
s_{1}=\frac{s_{0}\rho }{g}\qquad \qquad \text{and}\qquad \qquad s_{2}=\frac{
s_{0}}{a^{3\gamma }g},  \label{s2}
\end{equation}
respectively.

These results can be used to find exact solutions for $a(t)$ in universes
with two interacting fluids, and non-zero spatial curvature.

\end{document}